\documentclass[final,3p,times,twocolumn]{elsarticle}
\usepackage{graphicx}
\usepackage{amssymb}
\usepackage{amsthm}
\usepackage{lineno}

\journal{Nuclear Instruments and Methods, Section B}

\begin{document}

\begin{frontmatter}


\title{Bias and synergy in the self-consistent approach of data analysis of ion beam techniques}



\author{T. F. Silva\fnref{myfootnote}}
\fntext[myfootnote]{Corresponding author. e-mail: tfsilva@if.usp.br}
\author{C.L. Rodrigues}
\author{M.H. Tabacniks}
\address{Instituto de Física da Universidade de S\~ao Paulo, Rua do matão, trav. R 187,\\ 05508-090 S\~ao Paulo, Brazil.}
\author{U. von Toussaint}
\author{M. Mayer}
\address{Max-Planck-Institut f\"ur Plasmaphysik, Boltzmannstr. 2, D-85748 Garching, Germany.}

\begin{abstract}
Using multiple ion beam analysis measurements or techniques combined with self-consistent data processing generally allows for extracting more (or more accurate) information than processing data from single measurements separately. Solving ambiguities, improving the final depth resolution, defining constraints and extending applicability are the main strengths of the data-fusion approach, which essentially consists in formulating a multi-objective minimization problem that can be tackled by the adoption of the weighted-sum method. A simulation study is reported in order to evaluate the systematic error inserted in the analysis by the choice of a specific objective function, or even by the weights or normalization adopted in the weighted-sum method. We demonstrate that the bias of the analyzed objective functions asymptotically converges to the true value for better statistics and that the measurement can be ranked by its information content, where some combinations of measurements better constrain the multi-objective optimization.
\end{abstract}

\begin{keyword}
Self-consistent analysis \sep Ion beam analysis \sep Systematic errors \sep Objective function \sep Data fusion


\end{keyword}

\end{frontmatter}

%\linenumbers

\section{Introduction}
\label{intro}

A data-fusion approach for data analysis of spectrometry information obtained using Ion Beam Analysis (IBA) techniques is a powerful tool for improved material characterization, providing more reliability and increasing the quality of information extracted from these measurements \cite{totaliba}. It consists of applying as many IBA measurements as necessary on the same sample and then combining all the information in a unique sample model that describes all experimental data within some statistical significance. Usually, this model is found by an optimization algorithm in a computer program, given some combined objective function. We cite two codes that utilize this approach in data processing: WiNDF (hereby named NDF) and MultiSIMNRA.

%Usually, this model is found by an optimization algorithm in a computer program, given some combined objective function. Currently, only two GUI-based computer programs widely adopted by the IBA community enable a data-fusion approach to process data. The first to provide this features was WinNDF (hereby named simply as NDF), which uses DataFurnance as calculation engine \cite{Barradas97, NDFsimanneling}.  Created in the 90s, this code can handle different techniques, including Rutherford Backscattering Spectrometry (RBS), Nuclear Reaction Analysis (NRA), Elastic Recoil Detection (ERD), Elastic Backscattering Spectrometry (EBS), Particle Induced X-ray Emission (PIXE), among others \cite{totaliba, NDF_newtool}. The second to provide this feature is MultiSIMNRA \cite{multisimnra,multisimnra_new, NDF_externalbeam} whose first release occurred in 2015. Despite being new, MultiSIMNRA already offers many interesting features, conveniently organized into a user-friendly environment. It relies on the physical simulations provided by the widely adopted SIMNRA software \cite{simnra, simnra7}, being currently able to handle data from RBS, NRA, EBS, and ERD, Medium Energy Ion Scattering (MEIS). SIMNRA also provides calculations for Particle Induced Gama-ray Emission(PIGE), which is not yet supported by MultiSIMNRA. 

Created in the 90s, NDF uses DataFurnance as a calculation engine \cite{Barradas97, NDFsimanneling}. It can handle different techniques, including Rutherford Backscattering Spectrometry (RBS), Nuclear Reaction Analysis (NRA), Elastic Recoil Detection (ERD), Elastic Backscattering Spectrometry (EBS), Particle Induced X-ray Emission (PIXE), among others \cite{totaliba, NDF_newtool, NDF_externalbeam}. MultiSIMNRA \cite{multisimnra,multisimnra_new} had its first release in 2015. Despite being new, MultiSIMNRA already offers many interesting features, conveniently organized into a user-friendly environment. It relies on the physical simulations provided by the widely adopted SIMNRA software \cite{simnra, simnra7}, being currently able to handle data from RBS, NRA, EBS, and ERD, Medium Energy Ion Scattering (MEIS). SIMNRA also provides calculations for Particle Induced Gama-ray Emission(PIGE), which MultiSIMNRA does not yet support. 

The strength of the joint processing of IBA data lies in the synergy that occurs when combining the individual information contained in the different measurements. Butler \cite{butler} introduced the concept of using chemical or thermodynamic information in the analysis, as an alternative to additional measurements, in an attempt to constrain the solution of an ambiguous RBS measurement. In this sense, both the combination of measurements or the use of prior information are, in principle, two possible ways to improve the material characterization by means of a combined processing of the data \cite{multisimnra_new}.

More recently, Jeynes \cite{primarystd} has shown that RBS is a primary direct reference method with an unrivalled traceable accuracy for certain thin film measurements. This was demonstrated at nearly 1\% both in a multi-laboratory test \cite{jeynes_ac} and in a longitudinal test over several years in one lab \cite{Colaux_analyst}. Jeynes et al. \cite{totaliba} claimed that the data-fusion approach inherits the accuracy of the most accurate measurement in the dataset, allowing the absolute accuracy of RBS to be inherited by any synergistic analysis: this was demonstrated in detail by Total-IBA of a known glass \cite{jeynes2020_accuracy}. This is indeed reasonable if one thinks that an accurate measurement constraints better the solution space during the optimization process, being the major constraint also in the calculation of statistically acceptable solutions in the uncertainty evaluation (see section 2.2 in \cite{multisimnra_new}). In this sense, this assumption seems to be correct.

However, one important consequence of this assumption is that, given a certain set of measurements, possibly there are some other new measurements that can be performed and added in the analysis, which can improve the final accuracy. On the other side, there are other measurements that cannot succeed in this task of improving the final accuracy significantly, thus, are not worth to be performed. This is simply because adding a measurement in joint data processing can be considered as adding new constraints to the optimization algorithm, and there are measurements that constraint the parameters more strongly, and others that do not. Thus, we can say that some measurements combine synergistically, and others do not.

On top of that discussion, there is also the problem of bias (a systematic tendency which causes differences between the final result and the true value) which may be introduced by the choice of the likelihood function or by deficiencies of the forward model. On first glance the choice of the likelihood function appears to be straightforward for most of the ion-beam based methods: Individual events are being registered and the underlying physics (rare, independent scattering events) thus implies a Poissonian likelihood with a expected number of events $\lambda$ which depends on the analysed sample (with sample parameters $\theta_{s}$) and the diagnostic settings (experimental parameters like beam energy, projectile species, detector solid angle, sensitivity, energy resolution etc.), here summarized by $\theta_{d}$. Then the probability to observe $c$ counts is given by

\begin{equation}
	p_{P}\left(c\mid\theta, I\right) = \frac{\lambda\left(\theta\right)^{c}}{c!}\exp\left(-\lambda\left(\theta\right)\right),
\end{equation}	
with $\theta$ denoting the union of sample and diagnostic parameters: $\theta=\left\{\theta_{s},\theta_{d}\right\}$. For this likelihood it can be shown \cite{frieden91} that the estimation of the parameters is unbiased, i.e. the estimation converges to the correct parameter values with increasing number of data and that the theoretical optimum of the estimation accuracy, i.e. the Cramér-Rao Bound (CRB) is achieved. 

However, the likelihoods which are actually used are different and involve several intermediate approximations for a number of reasons. In an almost generic first step the Poissonian likelihood is approximated by a Gaussian likelihood, which holds with good accuracy for a sufficiently large number of counts. In a second approximation step the variance of the Gaussian likelihood is set as $c$, i.e. based on the actual observed number of counts

%\begin{figure*}
\begin{equation}
p_{G}\left(c\mid\theta, I\right) = \frac{1}{\sqrt{2\pi c}}\exp\left(-\frac{1}{2}\left(\frac{c-\lambda\left(\theta\right)}{\sqrt{c}}\right)^{2}\right),
\end{equation}	
%\end{figure*}
using $p_{G}\left(c\mid\theta, I\right) = N\left(\lambda\left(\theta\right),\sqrt{c}\right)$ instead of the mathematically correct $N\left(\lambda\left(\theta\right),\sqrt{\lambda\left(\theta\right)}\right)$. The justification for this second approximation rests on the improved numerical stability of the optimization. In the early stages of the optimization the model may yield values which are significantly different from the observed data and this discrepency is magnified by the simultaneously deviating value of the uncertainty, resulting in numerical overflow or convergence failure. In practice there is a second complication: Most measurements are affected by some background signal which needs to be accounted for. A proper statistical handling of this matter turns out to be surprisingly challenging because the difference of two Poisson distributed random variables is not longer described by a Poisson distribution but instead follows a Skellam distribution \cite{udo}. This is quite different from Gaussian random variables where their sum and their difference are again described by a Gaussian probability distribution. Also a Bayesian approach for a proper handling of the background subtraction yields a non-standard likelihood \cite{gregory}. Another remarkable example where these assumptions for the likelihood functions fail is in the case of x-rays spectroscopies (like PIXE and x-ray fluorescence) as reported in \cite{NDF_externalbeam, PappMaxwell}.

In addition there is a third reason why likelihoods used in data-fusion approaches are adjusted. The arguments about being asymptotically unbiased do hold only under the assumption that the forward model, i.e. the model relating the parameters $\theta$ and the expected number of counts $\lambda\left(\theta\right)$ is perfect. Unfortunately, although the models used in NDF and SIMNRA are continously improved there are inevitible approximations of the underlying scattering and detection process. These small deviations are often of no concern, especially if only a single diagnostic is being used. The problem commonly becomes apparent when diagnostics of very different count rates are jointly evaluated. Then a small model inaccuracy in one diagnostic can completely dominate annother diagnostic. The prototypical example is the combination of data from a forward scattering experiment with conventional RBS-data.  Tiny inaccuracies of the multiple-scattering modelling in the forward direction together with a large number of counts in this experiment yield a most likely result from the joint evaluation which are incompatible with information of the RBS measurement alone: the RBS contribution has been overwhelmed. For that reason sometimes the statistical weight of the individual measurements is being 'adjusted' - which may allow an otherwise impossible joint fit of different diagnostics but can also introduce a bias of unknown extend.

Therefore, this paper deals with these two important aspects of the data-fusion approach of processing IBA data: bias and synergy. Both issues directly impact the final accuracy of the result: while bias introduces systematic errors, the synergy obtained by the combination of different measurements constrains the result more strongly, thus reducing the uncertainties. Therefore, in this study, we aimed to better understand the uncertainties associated with the simultaneous processing of multiple data, and the influence of the choice in the objective function on the final accuracy.

\section{Methods}

We designed a simulation exercise in order to evaluate both, the systematic errors introduced on the final result by the bias of the objective functions, and the final accuracy when combining different measurements. Performing this study through simulations is justified because we aim at the evaluation of systematic errors induced only by the objective functions, while the analysis of experimental data is affected by systematic errors originating from different sources, such as the physics models \cite{mayer_possibilities,rauhala,summary_intercomparison,intercomparison} or the fundamental databases (e.g. stopping forces \cite{judging} and cross-sections databases \cite{abriola}). Another reason is: since we want to evaluate systematic errors and their uncertainties, we need to compare the optimum values of the objective function with true values, and this is only possible in simulations. Similar approaches have been used before, like in \cite{NDF_newtool, butler, Barradas_bayesian}.

%% novo texto %%
Another advantage of performing this study with simulations is direct access to the minimum value of the objective functions. Since we know the parameter's actual value, we also know what region of the solution space to explore. Thus, we calculate the values of all objective functions in a fine mesh within a hyper-volume of the solution space, which directly gives us the point of minimum. Doing so requires a long computational time but is essential to avoid issues related to the optimization algorithm or its different implementations. As a consequence, we can compare the objective functions only.

\subsection{The simulation exercise}

For the simulation exercise, we defined a sample consisting of 130 nm thin film of SiO$_2$ with 10\% H content deposited on top of an amorphous Silicon substrate. Then, using simulations provided by SIMNRA for different analysis conditions with Poisson noise added, we generated spectral data that played the role of experimental data. 

%Fig. \ref{fig:setup} shows the idealized setup to perform the calculations.
Two detectors were assumed: one in a backscattering geometry located at 170$^\circ$ scattering angle (referred to the incident beam direction); and another detector placed in a forward geometry located at 30$^\circ$ scattering angle. The solid angles of both detectors were assumed to be 1 msr. No electronics effects other than energy resolution of 12 keV (such as pile-up or dead-time) was added into the simulations.

%\begin{figure}[h]
%\label{fig:setup}
%\centering\includegraphics[width=0.8\linewidth]{fig1.png}
%\caption{Idealized setup to perform calculations.}
%\end{figure}

\begin{figure*}[h]
\label{fig:setup}
\centering\includegraphics[width=0.9\linewidth]{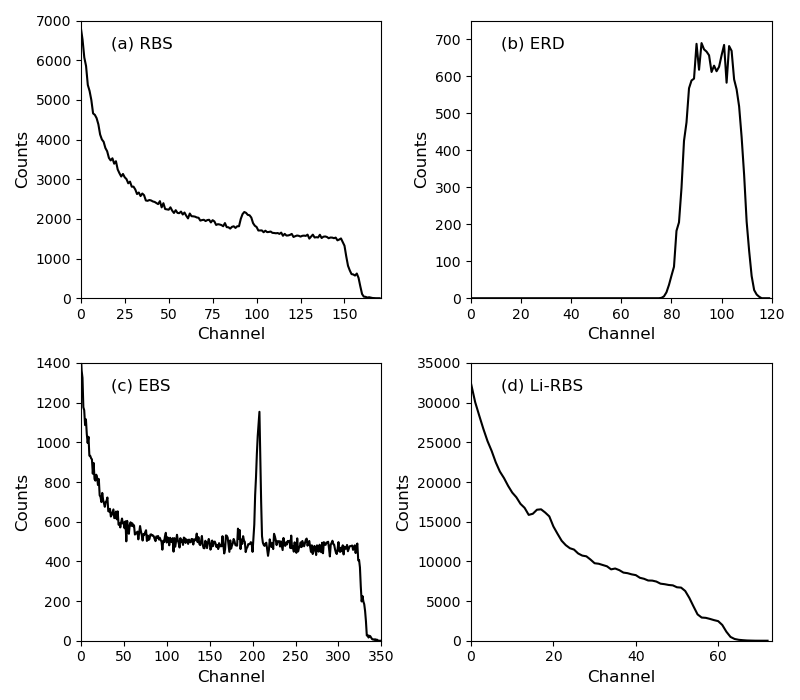}
\caption{Typical spectra for the four idealized experiments with Poisson noise added. Examples show simulations performed using 10 $\mu$C integrated charge.}
\end{figure*}

Thus, in this geometrical configuration, the detector placed at forward geometry was used to measure the H content of our hypothetical sample by ERD using He ions as a probe (a foil was placed in front of the detector to block scatered He ions), and the detector placed at backscattering geometry was used to measure Si and O content on it. For the latter, three configurations for ion and energy were adopted: one for He RBS with 1.5 MeV beam (the same energy as adopted for ERD, thus performed simultaneously), one for EBS with 3.04 MeV He beam (to take advantage of the resonant cross-section for O to enhance its signal in the spectra), and a last experiment for RBS with 1.0 MeV Li beam. This beam was assumed to provide an improved depth resolution due to its higher stopping force. In a real measurement this effect would be somewhat smaller than in our simulations (where we used identical detector energy resolutions for He and Li) due to the deterioration of the detector energy resolution for Li compared to He. In this sense, three virtual experiments were performed and are summarized in table \ref{table}.

We also aim to study the bias introduced by the objective function in the full analysis and uncertainties estimates of the RBS+ERD experiment. Each measurement has some level of bias given by its level of noise, and the bias of the combined result is what we want to evaluate here. Therefore, we want to assess the role of the integrated charge (statistical significance of the spectra) on that bias. After that, we want to study which measurement adds more information to the analysis, whether it is the EBS measurement by the enhanced oxygen signal or the Li-RBS with better depth-resolution, given a fixed integrated charge (10 $\mu$C). To obtain the predictions using each objective function we used the method described above.

%\begin{table*}[ht!]
%\centering
%\begin{tabular}{ccccc}
%\hline
%                   & \textbf{Incident} & \textbf{Integrated} & \textbf{Scattering} & \\
%\textbf{Technique} & \textbf{beam}     & \textbf{charge}     & \textbf{angle}      & \textbf{Goal}\\
%\hline
%RBS+ERD & 1.5 MeV & 5, 10  & 170 & Full \\
%& He & and 20 $\mu$C & 30 & characterization \\
%\hline
%EBS & 3.04 MeV & 10 $\mu$C & 170 & O signal \\
% & He & & & enhancement \\
%\hline
%Li-RBS & 1.0 MeV& 10 $\mu$C & 170 & Improvement in \\
% & Li & & & depth-resolution \\
%\hline
%\end{tabular}
%\caption{Summary of experiments.}
%\label{table}
%\end{table*}

\begin{table*}[ht!]
\centering
\caption{Summary of experiments.}
\begin{tabular}{ccccc}
\hline

                   & \textbf{Incident} & \textbf{Integrated} & \textbf{Scattering} & \\
\textbf{Technique} & \textbf{beam}     & \textbf{charge}     & \textbf{angle}      & \textbf{Goal}\\

%\textbf{Technique}   & \textbf{Incident beam} & \textbf{Integrated charge} & \textbf{Scattering angle} & Goal \\
\hline
RBS+ERD & 1.5 MeV He & 5, 10, 20 $\mu$C  & 170$^\circ$, 30$^\circ$ & Full characterization \\
%\hline
EBS & 3.04 MeV He & 10 $\mu$C & 170$^\circ$ & Enhance O signal \\
%\hline
Li-RBS & 1.0 MeV Li & 10 $\mu$C & 170$^\circ$ & Improve depth-resolution \\
\hline
\end{tabular}
\label{table}
\end{table*}

\subsection{Tested objective functions}

We considered three objective functions in our tests. The simplest form on the list was the sum of the $\chi^2$ for the different spectra. 

\begin{equation}
\label{std_chi2}
F_{\chi^2} = \sum_{\textrm{Spectra}} \left [ \sum_{\textrm{Channels}} \left( \frac{c_m-c_i}{\sigma_i} \right)^2 \right ]
\end{equation}
where $c_m$ is the number of counts in each channel calculated using the forward model (simulation) and the $c_i$ is the number of counts on each channel for the experimental spectra.  $\sigma_i$  is the estimated uncertainty of $c_i$ (assuming Poisson distribution it is equal to $c_i^{1/2}$ or equal to one in case $c_i=0$).

The second function was the MultiSIMNRA objective function, which is based on the weighted-sum method for multi-objective optimization \cite{multisimnra, multisimnra_new}. It scales the individual $\chi^2$ spectrum by its expected value so they have the same expected minimum value, therefore the same relative importance for the optimization algorithm \cite{survey}. 

\begin{equation}
\label{MS_chi2}
F_{MS} = \frac{1}{S} \sum_{\textrm{Spectra}} \left [ \frac{1}{DoF} \sum_{\textrm{Channels}} \left( \frac{c_m-c_i}{\sigma_i} \right)^2 \right ]
\end{equation}
where DoF is the number of degrees-of-freedom of the fit and S is the total number of spectra.

The third tested objective function was the NDF objective function. This is not based on the standard $\chi^2$, but it is based on the sum of squared differences of the simulated and experimental spectra. The normalization factor, in this case, is the area of each spectrum to the 1.5 power.

\begin{equation}
\label{NDF_chi2}
F_{NDF} = \sum_{\textrm{Spectra}} \left [ \frac{1}{A_j^{1.5}}\sum_{\textrm{Channels}} \left( c_m-c_i \right)^2 \right ]
\end{equation}

In fact, the area of the spectra is the expected value for the sum of squared differences (assuming Poisson distribution). However, according to the authors, the 1.5 power on the normalization is inserted ad-hoc for performance purposes \cite{unambiguous}. The original NDF objective function also has a term that penalizes the optimization algorithm in case it increases the number of parameters in the fit \cite{NDFsimanneling}. But this term was not inserted here since we kept the number of fitting parameters always fixed.

Other objective functions may be available in NDF, mainly for the Bayesian inference method of uncertainty estimation \cite{Nuno}. We refer to eq. \ref{NDF_chi2} as an alternative example, and as the only version published until now for the NDF's objective function.

\section{Results}

\subsection{Influence of the counting statistics}

\begin{figure*}[htb!]
\centering\includegraphics[width=1.0\linewidth]{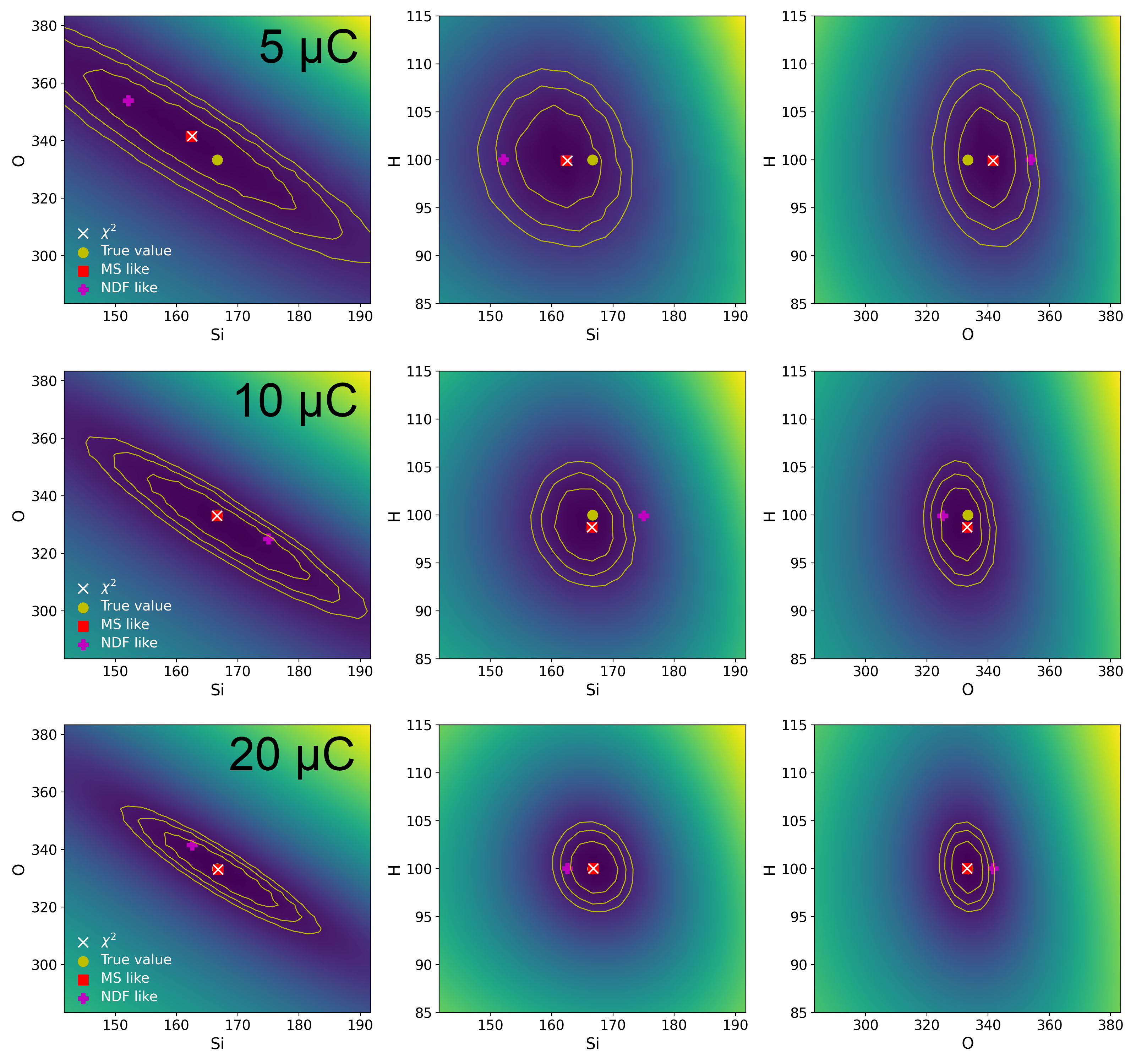}
\caption{Heat-map (in log scale) for the $\chi^2$ objective function and the optimum points of the three different objective functions for the simulated case of 5 $\mu$C (upper line), 10 $\mu$C (middle line) and 20 $\mu$C (lower line) integrated charge. Axis units are $1\times10^{15}$ at./cm$^2$. The true value used to generate the data is also shown to illustrate the bias introduced by the noisy data into the objective functions. The continuous curve denotes one standard deviation defined by the $\chi^2$ distribution. Where the true value is not observed, it lies beneath the $\chi^2$ and the MS-like point. }
\label{stattest}
\end{figure*}

Here we evaluate the impact of the counting statistics on the evaluation of the RBS+ERD data with the different objective functions, which the major influence is constraining the bias of the objective function. Increasing the integrated charge of the spectra makes the objective functions less susceptible to the effects of the Poisson noise. It can be observed in Fig. \ref{stattest} that all the minima of the objective functions converge asymptotically to the true value with increasing integrated charge. The effects on the bias introduced by the noise are apparently more critical for NDF-like objective functions, since the optimal value predicted by this function lies outside of the confidence interval for the lowest tested value of the integrated charge, as can be observed in Fig. \ref{stattest} (upper).

It is worth to mention that the positions of functions minima changes from one simulation to another. The only point that does not change its position is the true value (yellow dots). All others are noise dependent, thus each time random noise is added, the position of the minimum changes. Results shown here are representative to many consecutive simulations, and illustrate the author's arguments.

The confidence region's elliptical shape and its inclination indicate the negative correlation between the evaluations for the Si and O amounts. It is because the measured energy loss of the layer is given essentially independently of the elements' amounts ratio (if there is less Si there must be more O to result in the same energy loss). However, the information on the Si/O ratio is given by the height of the signals in the spectra, thus highly susceptible to counting statistics (not all pairs of values of Si and O amounts fit the spectra heights). These two constraints together result in this elliptical shape. When the counting statistics grow, more defined is the measurement of the energy loss, and better is the Si/O measurement, reflected in the shrinkage of the confidence region.

\subsection{Combination with EBS}

In principle, the EBS measurement is intended to take advantage of the resonant cross-section that occurs at 3.038 MeV for the $^{16}$O($\alpha$,$\alpha$)$^{16}$O reaction \cite{sigmacalc, Colaux_o16}. The resonance enhances the $^{16}$O signal in the spectra, thus increasing the counting statistics in the oxygen peak. However, increasing the energy also reduces the effective stopping power and as a consequence reduces the depth-resolution.

\begin{figure*}[ht!]
\centering\includegraphics[width=1.0\linewidth]{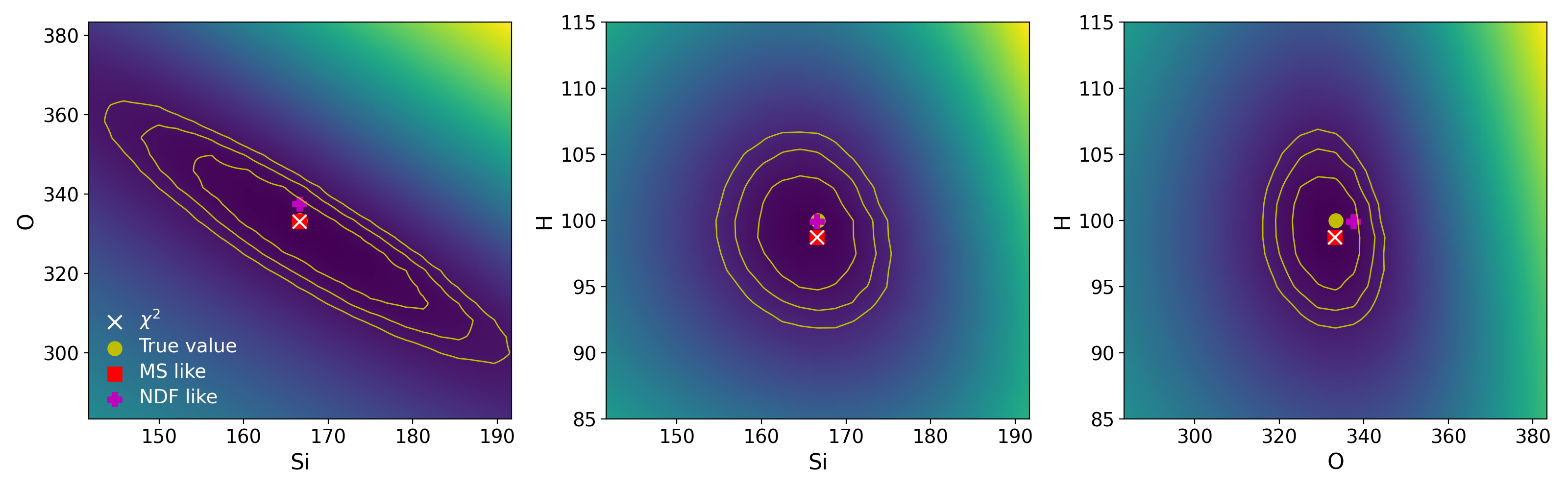}
\caption{Heat-map (in log scale) for the $\chi^2$ objective function and the optimum points of the three different objective functions for the simulated case of 10 $\mu$C integrated charge and EBS measurements combined. See Fig. 2 for captions for axis, points and continuous curve. Where the true value is not observed, it lies beneath the $\chi^2$ and the MS-like point.}
\label{ebs10}
\end{figure*}

The simulations indicate that, instead of providing steeper constraints to the objective function, it contributes very little to the final result since the individual contribution to the objective function is broader in the case of EBS. This apparently is a direct consequence of the loss of depth-resolution. This is observed by no relevant difference between Fig. \ref{stattest} (middle) and Fig. \ref{ebs10}.

\subsection{Combination with Li-RBS}

Since the worst depth-resolution resulted in a broader objective function, the Li-RBS measurement is intended to improve this situation by taking advantage of a higher stopping forces for the heavier ion. It is worth to point out that these simulation exercises were performed despite the less accurate database of stopping forces to Li. In fact, in an actual analysis, this should be included as a source of systematic error in the uncertainty budget. Here, however, the database is assumed as accurate since we want to study the effects of the insertion of a measurement with a better depth resolution as a constraint to the objective function.

\begin{figure*}[ht!]
\centering\includegraphics[width=1.0\linewidth]{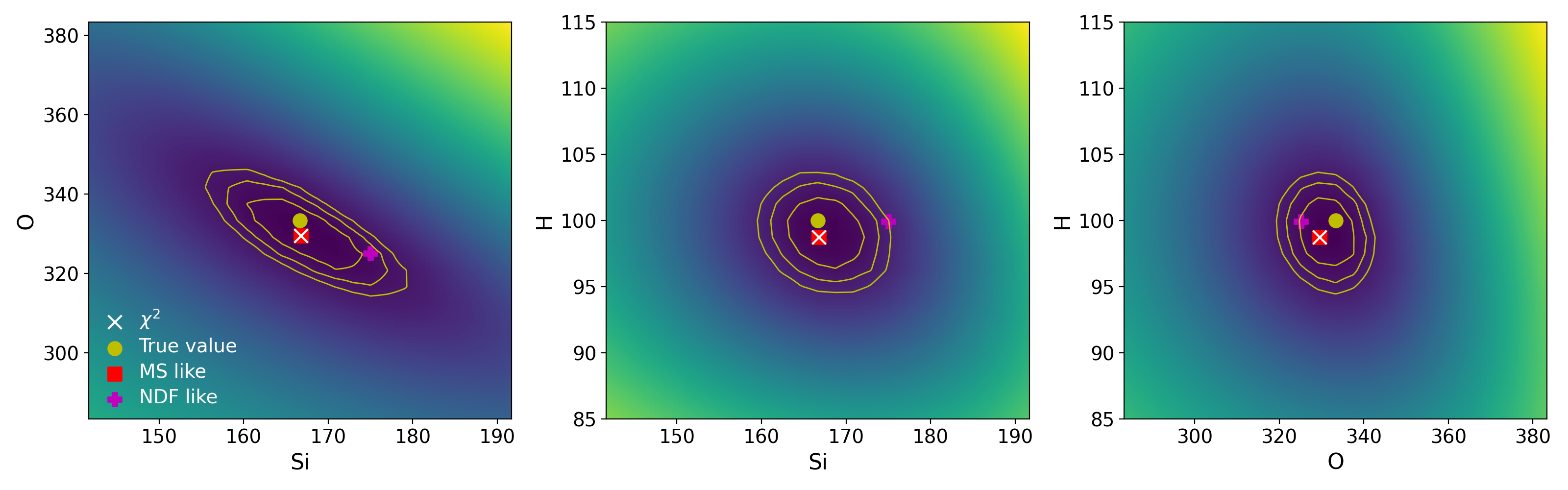}
\caption{Heat-map (in log scale) for the $\chi^2$ objective function and the optimum points of the three different objective functions for the simulated case of 10 $\mu$C integrated charge and Li-RBS measurements combined. See Fig. 2 for captions for axis, points and continuous curve. Where the true value is not observed, it lies beneath the $\chi^2$ and the MS-like point.}
\label{lirbs10}
\end{figure*}

Indeed, all resulting objective functions including the Li-RBS measurement are steeper and resulted in a more constrained fit. This is observed comparing Fig. \ref{stattest} (middle) and Fig. \ref{lirbs10}.

\section{Discussion}
With the simulated data, the gain of information was clearly observed when inserting the Li-RBS into the optimizations by the shrinkage of the confidence region, which is the region delimited by the uncertainty ellipse. On the other hand, no gain was observed when inserting the EBS analysis into the optimization due to the apparent sameness of the confidence region.

A possible explanation for this can be obtained in the Bayesian framework \cite{mackay,udo}. The Bayes theorem states a relationship between the probability distribution function (pdf) for the parameters ($\theta$) prior the inclusion of a new experiment $p(\theta|I)$, with the final state of the pdf in the light of a new experiment $p(\theta|D,I)$. This relationship depends on the likelihood function of the new measurement $p(D|\theta,I)$, and a normalization term called evidence, $p(D|I)$ \cite{udo_revmod}:

\begin{equation}
\label{bayes}
p(\theta|D,I) = \frac{p(\theta|I) \cdot p(D|\theta,I)}{p(D|I)}
\end{equation}

We can visualize what happens with the pdf when updated with new experimental data by assuming the evidence as a constant, and calculating the product of the prior pdf (the likelihood function of the previous experiment) with the likelihood function of the new measurement, i.e. the nominator in Bayes’ theorem. The heat maps presented in fig. 7 show this. For practical reasons, we show data only for Si and O parameters, however, similar maps can be produced using any combination of Si or O with the H parameter.

\begin{figure*}[h!]
\centering\includegraphics[width=1.0\linewidth]{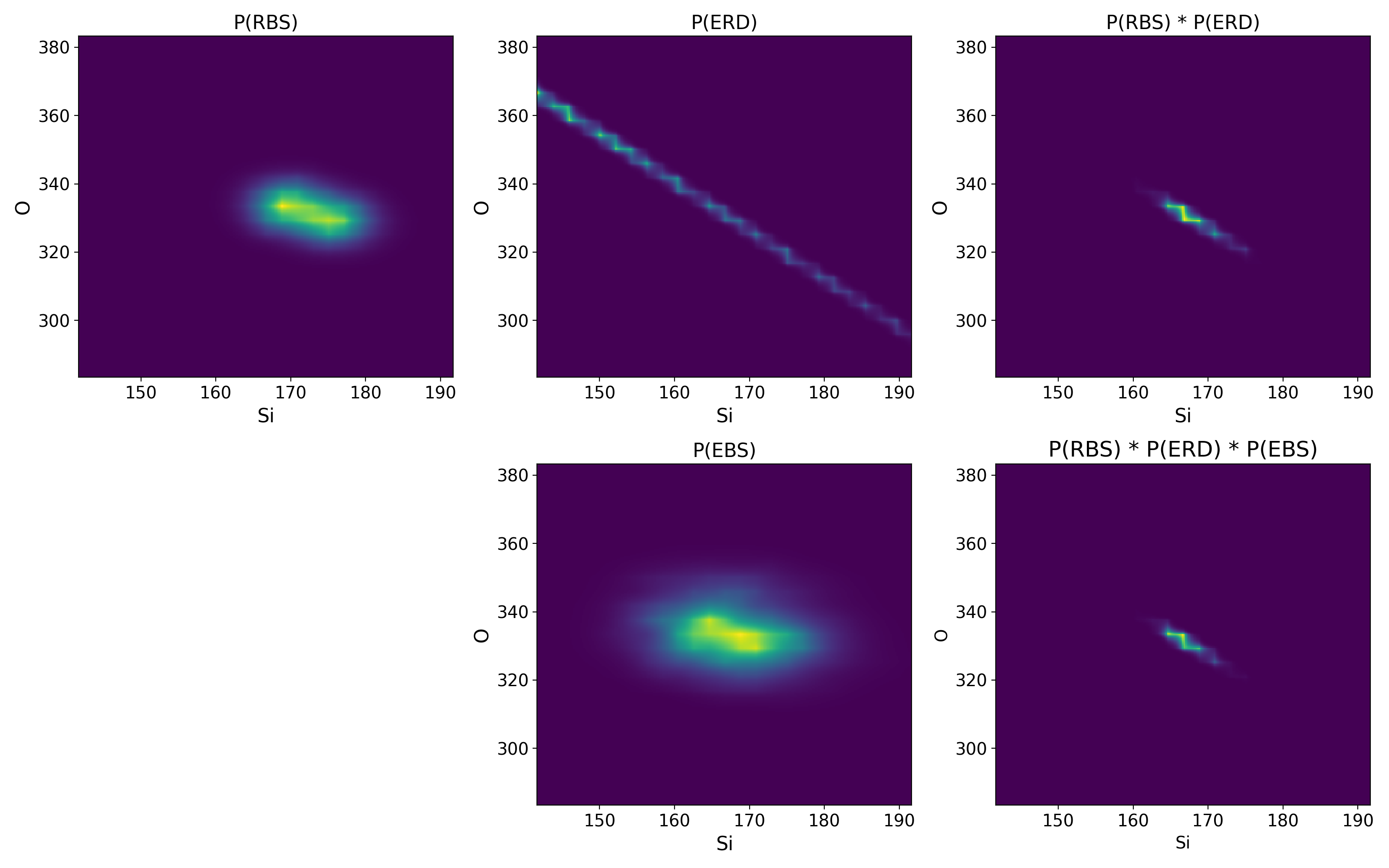}
\caption{Probability density functions resulting from the combination of RBS+ERD+EBS. In light of the Bayes theorem, the EBS measurement does not provide additional information since the pdf is broader than the prior given by the RBS and ERD combined. Axis units are $1\times10^{15}$ at./cm$^2$.}
\label{probebs}
\end{figure*}

The upper left figure in the panel of fig. \ref{probebs} shows the Si and O pdf given the RBS measurement. The upper middle figure shows the same but for the ERD measurement. Note that the ERD measurement only contains direct information for the H, and indirect information on the total amount of Si plus O, roughly given by the width of the H peak. The product of both pdfs results in the upper right figure, being the pdf in the light of the combination of the data contained in the RBS and the ERD data together. Finally, the pdf for the EBS measurement is presented in the lower middle figure, and the pdf in the light of the combination of the three measurements is presented in the lower right figure. 

\begin{figure*}[h!]
\centering\includegraphics[width=1.0\linewidth]{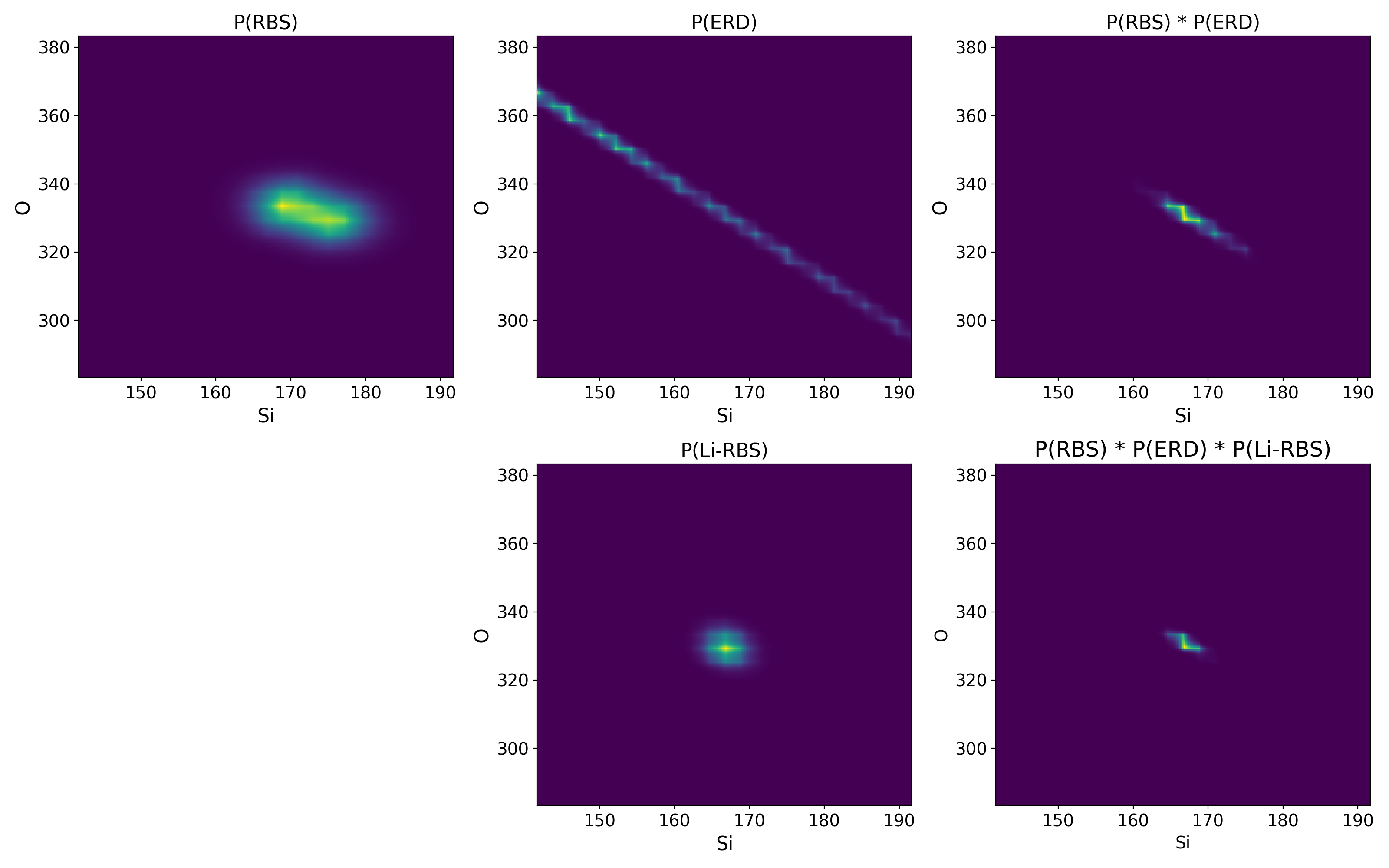}
\caption{Probability density functions resulting from the combination of RBS+ERD+Li-RBS. In light of the Bayes theorem, the Li-RBS measurement provides additional information since the pdf is narrower than the prior given by the RBS and ERD combined. Axis units are $1\times10^{15}$ at./cm$^2$.}
\label{problirbs}
\end{figure*}

Figure \ref{problirbs} tells a different story. While the EBS measurement presents a likelihood function that is broader than the prior (the pdf obtained with the combination of RBS and ERD), the likelihood function of the Li-RBS measurement is narrower (see the figure in the lower middle in the panel of fig. \ref{problirbs}). In this sense, the Bayes theorem results is a more restricted pdf, indicating the gain of information.

Concerning the bias of the objective functions. This can also be analyzed in the Bayesian framework. Since the new pdf in light of the new experiment gets less broad, the solution space gets more restricted, thus the optimal prediction deviates less from the true value, therefore converging to the region of maximum probability. This is an important result that demonstrates the synergy as a method to control the bias of the objective functions.

%\section{Outlook}

%The preceding results clearly demonstrate that different measurements result in differing probability distributions of the parameters. Typically more localized pdfs are preferred, i.e., the ones with lower entropy. It indicates that the expected entropy reduction caused by a measurement (or a sequence of measurements) can provide guidance to assess the value of another measurement or experimental technique. It thus opens the pathway towards quantitative experimental design \cite{chaloner,udo_revmod}. For ion beam applications, a case study on deuterium depth profiling focusing on NRA and optimal selection of beam energies has been given in \cite{udo_optimizing}.  It appears that a systematic study of the gains achievable by combining different diagnostic tools holds great promise and can result in significant efficiency gains. 

\subsection{Gain of information}

The shrinkage of the pdf observed in the figs. \ref{probebs} and \ref{problirbs} is a direct consequence of the gain of information provided by the IBA techniques. A narrow distribution reflects less uncertainty on the parameters, thus a state of more information.

The theory provides a quantitative scale for the information gain by the Kullback-Leibler divergence ($D_{\textrm{KL}}$) that measures the relative entropy between two pdfs. Here it expresses the difference in the state of information if the pdf in light of the new data is used instead of the prior pdf. A standard unit for information gain is the $bits$. 

\begin{equation}
\label{KLdivergence}
D_{\textrm{KL}}(P|Q) = \int p(\theta|D,I) \log_2 \left( \frac{p(\theta|D,I)}{p(\theta|I)} \right) d \theta
\end{equation}

The table \ref{information_gain} expresses the information gained for the specific case of this sample of each technique alone and when combined. The estimates for the techniques alone take as a reference a neutral prior (representation of ignorance), represented by a uniform pdf that extends from zero up to twice the true value in the three-axis variables (Si, O, and H).

\begin{table*}[h!]
\label{information_gain}
\centering
\caption{Information gained on different stages of the joint approach of data analysis. Values calculated using the Kullback-Leibler divergence.}
\begin{tabular}{ccc}
\hline
Prior & Posterior & Information gain ($bits$)  \\
\hline
Neutral & RBS & 12.8  \\
Neutral & ERD & 16.8  \\
Neutral & LiRBS & 15.0 \\
Neutral & EBS & 11.5  \\
RBS & RBS+ERD & 6.5 \\
RBS+ERD & RBS+ERD+LiRBS & 0.6 \\ 
RBS+ERD & RBS+ERD+EBS & 0.1 \\ 
%Ignorance & RBS+ERD+LiRBS & 20.4 \\
%Ignorance & RBS+ERD+EBS & 19.1 \\
\hline
\end{tabular}
\end{table*}

One can observe the technique alone that presents the highest information gain starting at the neutral prior is the ERD, followed by RBS with lithium probe, while the one with a minor gain is EBS. However, it is essential to notice that this value accounts not just for the increment in oxygen sensitivity enabled by the resonant cross-section but also considers the reduced depth resolution due to the higher energy of the helium probe.

The ERD case is interesting since it increases mutual information, i.e., how much one variable tells us about another. It happens because the width of the H peak introduces a strong constraint between the Si and O amounts.

Additionally, the information gained when combining ERD (posterior) with RBS (prior) is lower than the direct sum of the information gain of the separate techniques, indicating information does not add linearly in this case. It happens because part of the information on both measurements is redundant.

%Additionally, combining ERD with RBS leads to an information gain lower than that gained with ERD alone. It happens because part of the information on both measurements is redundant. Hence the information does not add linearly in this case. 

Finally, combining Li-RBS (posterior) with the RBS+ERD information state (prior) results in a six-fold information gain compared to the case of combining EBS (posterior) with the same RBS+ERD information state (prior). This is a quantitative measurement of what was observed in figs. \ref{probebs} and \ref{problirbs}.

%Finally, combining RBS with lithium probe results in a six-fold information gain compared to combining EBS, both starting from the RBS plus ERD state. 

\section{Conclusions}

In the self-consistent approach of analysis of multiple measurements, the forward model takes certain parameters, like the description of the sample proposed by the optimization algorithm, and computes a simulated spectrum that can be compared to the experimental observations. The optimization algorithm uses an objective function as a measure of the goodness of the fit, providing information to the algorithm to adjust the parameters in the search for the optimal parameters. 

This search consists in exploring the solution space looking for the minimum of the objective function, which is considered as the optimal estimate to the true value. Deviations on that estimate are expected due to the susceptibility of the objective functions to noise. Here, we demonstrated that, even in conditions of low statistics, the objective function adopted in MultiSIMNRA is robust, presenting a low susceptibility to noise. The objective function adopted by NDF displayed a wider scatter around the true value for consecutive runs of the code, indicating some persistent sensitivity to noise even at higher values of integrated charge or in combination to other measurements. Another result is that all objective functions tested converged asymptotically to the true value as higher the counting statistics (or integrated charge).

Besides that, we also demonstrated that incorporating multiple measurements by the adoption of the weighted-sum method can result in a gain of information. This depends on the likelihood function of the new measurement when compared to the pdf prior to the new measurement. If the likelihood function of the new measurement is broader than the pdf representing the current status of information, then no significant gain of information is observed. Alternatively, if the likelihood function is narrower than the prior pdf, then gain of information occurs.

In fact, this can be interpreted as a confirmation that the consistent data-fusion approach inherits the accuracy of the most accurate measurement since this offers the most stringent constraint to the optimization algorithm. However, this also establishes that some possible measurements, when added to a pool of measurements processed self-consistently, may not result in a relevant gain of information, depending if their likelihood functions combine synergistically or not.

Besides, the preceding results clearly demonstrate that different measurements result in different probability distributions of the parameters. Typically more localized pdfs are preferred, i.e., the ones with lower entropy. It indicates that the expected entropy reduction caused by a measurement (or a sequence of measurements) can provide guidance to assess the value of another measurement or experimental technique. It thus opens the pathway towards quantitative experimental design \cite{chaloner,udo_revmod}. For ion beam applications, a case study on deuterium depth profiling focusing on NRA and optimal selection of beam energies has been given in \cite{udo_optimizing}.  It appears that a systematic study of the gains achievable by combining different diagnostic tools holds great promise and can result in significant efficiency gains. 

\section{Acknowledgement}

The authors thank the financial support given by CNPq-INCT-FNA (project number 464898/2014-5).



%% References


\bibliographystyle{model1-num-names}
\bibliography{references.bib}






% \end{thebibliography}

\end{document}